\documentclass{pasa}%

\usepackage{graphicx}
\usepackage{amsmath}
\usepackage{amssymb}

\title[SimSpin]{SimSpin - Constructing mock IFS kinematic data cubes}

\author[K.E. Harborne et al.]{K. E. Harborne$^{1,2}$, C. Power$^{1,2}$ and A. S. G. Robotham$^{1,2}$
\affil{$^1$International Centre for Radio Astronomy (ICRAR), M468, The University of Western Australia, 35 Stirling Highway, \\ 
Crawley, WA 6009, Australia\\}%
\affil{$^2$ARC Centre of Excellence for All Sky Astrophysics in 3 Dimensions (ASTRO 3D)}
}%

\jid{PASA}
\doi{10.1017/pas.\the\year.xxx}
\jyear{\the\year}

\usepackage{aas_macros}
\usepackage{hyperref} 
\hypersetup{colorlinks,citecolor=blue,linkcolor=blue,urlcolor=blue}
\usepackage{listings}
\usepackage{booktabs}
\newcommand{\gadget}[1]{\textsc{GADGET-2}#1} % typeset for Gadget2
\newcommand{\galic}[1]{\textsc{GalIC}#1} % typeset for GalIC
\newcommand{\simspin}[1]{\textsc{SimSpin}#1} % typeset for SimSpin

\newcommand{\eagle}[1]{\textsc{Eagle}#1}
\newcommand{\prospect}[1]{\textsc{ProSpect}#1}

\begin{document}

\begin{frontmatter}
\maketitle

\begin{abstract}
We present \simspin{}, a new, public, software framework for generating integral field spectroscopy (IFS) data cubes from $N$-body/hydrodynamical simulations of galaxies, which can be compared directly with observational datasets. \simspin{} provides a consistent method for studying a galaxy's stellar component. It can be used to explore how observationally inferred measurements of kinematics, such as the spin parameter $\lambda_R$, are impacted by the effects of, for example, inclination, seeing conditions, distance, etc. \simspin{} is written in {\small R} and has been designed to be highly modular, flexible, and extensible. It is already being used by the astrophysics community to generate IFS-like cubes and FITS files for direct comparison of simulations to observations. In this paper, we explain the conceptual framework of \simspin{}; how it is implemented in {\small R}; and we demonstrate \simspin{}'s current capabilities, providing as an example a brief investigation of how numerical resolution affects how reliably we can recover the intrinsic stellar kinematics of a simulated galaxy. 

\end{abstract}

\end{frontmatter}

\section{INTRODUCTION}
\label{sec:intro}

Over the last few decades, we have seen vast improvements in our understanding of galaxy evolution by combining photometric measurements of galaxies with observed, projected, stellar kinematics. While photometry provides clues about a galaxy's assembly history, the inclusion of kinematics has revealed a whole new perspective that highlights the imprints of accretion and merger events. These imprints can be quantified and connected to the mass and environment of a galaxy. This new perspective has opened up new avenues to investigate the drivers of galactic evolution \citep{Binney2005RotationRevisited, Emsellem2007TheGalaxies, Cappellari2011TheRelation, Cortese2016TheMorphology, vandeSande2017TheSurveys}. During the same period, advances in numerical simulations of galaxy formation and evolution have enabled comparable kinematic measurements to be made of simulated galaxies. This has provided a physically motivated framework to understand how galactic structure and kinematics are entwined, and to interpret the astrophysical significance of observed kinematic signatures \citep{Jesseit2009SpecificParameter, Naab2014TheRotators, Teklu2015ConnectingMorphology, Lagos2018QuantifyingGalaxies}.

As observations and simulations have grown in both scope and sophistication, the question of how to compare them in a faithful manner has become more important. The now well-established standard approach is to generate synthetic data products from theoretical datasets. This substantially reduces the inherent uncertainties in translating from an observed dataset to an estimate of the physical quantity of interest. The mock images produced can be passed through the same software tools that observers use to give consistent comparisons, and also allows for the incorporation of observational limitations, such as the effects of the atmosphere that can artificially distort the observed line-of-sight (LOS) velocities. This approach is already being pursued by the SAMI \citep[the Sydney-AAO Multi-object Integral field spectrograph survey;][]{Croom2012TheSpectrograph, Bryant2015TheSelection} and MaNGA \citep[Mapping Nearby Galaxies at Apache Point;][]{Bundy2015OverviewObservatory, Blanton2017SloanUniverse} teams \citep[see, for example,][]{Lagos2018TheGalaxies, Bassett2019ProspectsQuantities, Duckworth2020DecouplingSpin}. This alone suggests that a tool for creating such data products in a publicly accessible and repeatable way is advantageous for the community.

\medskip

Mock data products from large cosmological simulations such as \textsc{Millenium} \citep{Springel2005SimulationsQuasars} and \textsc{Illustris} \citep{Vogelsberger2014PropertiesSimulation} are available via corresponding ``observatories'' - the Millenium Run Observatory, MRObs \citep{Overzier2013TheLight} and the Illustris Simulation Observatory \citep{Torrey2015SyntheticSimulation}. Tools such as SISCO \citep[Simulating IFU Star Cluster Observations;][]{Bianchini2015UnderstandingObservations} have also been used to generate IFS observations of globular clusters for exploring kinematic signatures of intermediate-mass black holes \citep{DeVita2017ProspectsSpectroscopy}. However, similar complex data products for galaxy-scale models can be difficult and time-consuming to produce, and so often the focus has been on bulk physical properties \citep{Overzier2013TheLight}, such as the angular momentum and structure of galaxies \citep{Genel2015GalacticSequence, Teklu2015ConnectingMorphology, Pedrosa2015AngularScenario}. 

This approach is no longer viable, however; not only does it limit the complexity of the observational data that can be compared to, but it also limits how these data can be used to benefit theoretical modeling. Mock observations not only assist our interpretation of observable kinematics, but also allow us to tune our sub-grid physics models within simulations. To understand whether our models of feedback in simulations are sensible, we need to be investigating the more detailed gas and stellar kinematics as well as comparing to the cutting edge HI \citep{Papastergis2017TestingDwarfs} and IFS surveys \citep{vandeSande2018TheSimulations}. Sub-grid recipes used in the latest galaxy formation simulations tend to be based on older stellar formation and feedback models; for example, \textsc{Illustris} \citep{Genel2014IntroducingTime}, and its successor \textsc{IllustrisTNG} \citep{Pillepich2018SimulatingModel}, models the star-forming inter-stellar medium (ISM) gas as an effective equation of state, first proposed by \cite{Springel2003CosmologicalFormation}. This approach is common in simulations where ISM structure is below the resolution limit of the model \citep{Ascasibar2002NumericalHistory, Few2012RAMSES-CH:Simulations}. However, as simulations drive towards higher resolutions and start to model, for example, the star-forming ISM in more detail, the kinds of comparisons required to verify the utility of these models must similarly become more sophisticated.

\medskip

These considerations have led us to develop \simspin{}, a framework to allow for a fair comparison of simulated and observed datasets. \simspin{} is a modular R-package that takes a particle model and creates a synthetic stellar-kinematic data cube from which we can generate mock flux, LOS velocity, and LOS velocity dispersion images using the specifications of any IFS. Observational effects, such as the resolution of the cube and distortions in the atmosphere, can be incorporated. From these images we can study the specific effects that observing has on the kinematic properties recovered, given that we have access to the intrinsic properties of the model under scrutiny. Each simulated galaxy can be analyzed many times from a range of projected distances and angles. 

\simspin{} is designed to be quick and repeatable, allowing a small number of models to produce a large number of observations. This is the first open-source package of its kind (registered with the Astrophysics Source Code Library \citep{Harborne2019SimSpin:Simulations}), and allows any astronomer to generate mock kinematic images for comparison with real observations. This code can work with simple N-body models, but also has facilities to incorporate simple stellar population (SSP) synthesis models for processing hydrodynamic simulations. All data products can be output in a FITS file format similar to what would be produced by an observation. Furthermore, it is written in a highly modular fashion that allows modifications and extensions to be easily added in the future, i.e. radiative transfer outputs, dust screens, telescope specifics and further kinematic data manipulation. 

The purpose of \simspin{} is to ease the communication between practical and theoretical astronomers and accelerate our progress in understanding how specific stellar kinematic features evolve over time. Here, we present the initial framework for creating synthetic IFS kinematic data cubes from simulated galaxies. In Section \ref{sec:method}, we briefly describe the methodology of the code and show how each function of the package can be implemented in Section \ref{sec:implementation}. Full astronomical examples can be found in Section \ref{sec:examples}. Finally, we discuss possible extensions of this work and give a summary in Section \ref{sec:further}.

\section{METHODOLOGY}
\label{sec:method}

The purpose of the \simspin{} package is to take a simulation of a galaxy and to produce a data cube corresponding to that which would be obtained if it had been observed using an IFS -
spatial information in projection with kinematic information along the line-of-sight. A kinematic data cube can be produced using the functions in this package, from which ``observables'' can be measured and compared to the true (i.e. intrinsic) kinematic properties of the simulation. 

In this section, we present the methodology chosen to achieve this in a consistent and repeatable way:

\begin{enumerate}
\item Understand the true kinematics of the model.
\item Construct the simulation particle data into an ``observable format'' and bin data into a 3D kinematic cube.
\item Convolve the data cube with a point spread function (PSF) in order to replicate the effects of the atmosphere.
\item Construct synthetic images from 3D mock data cube. 
\item Calculate observable properties from the images produced (i.e. measure the effective radius of the galaxy, calculate the observable spin parameter within a given radius, etc.)
\end{enumerate}

We will briefly address the approach to each of these matters in turn.

\subsection{Understanding intrinsic model properties}
\label{sec:method1}
It is necessary to understand the inherent nature of the galaxy model in question in order to assess how observation impacts kinematic measurements. Hence, \simspin{} provides a method for analysing the phase-space information of the particles within a simulation before constructing the mock observables. 

Particle-based simulations provide the user with the particle IDs, positions ($x$, $y$, $z$), velocities ($v_x$, $v_y$, $v_z$) and masses. These properties are used within \simspin{} to describe the intrinsic physical and kinematic profiles of the galaxy. To construct profiles, we take each particle and compute several additional properties. The particle phase space distribution is centered by subtracting the median in position and velocity space. The radial distribution of the physical properties can then be mapped. We add the spherical polar coordinates for each particle ($r$, $\theta$, $\phi$), the corresponding velocities ($v_r$, $v_{\theta}$, $v_{\phi}$) and components of the angular momentum ($J_x$, $J_y$, $J_z$). 

\begin{figure}[!ht]
\centering
\includegraphics[width=\columnwidth]{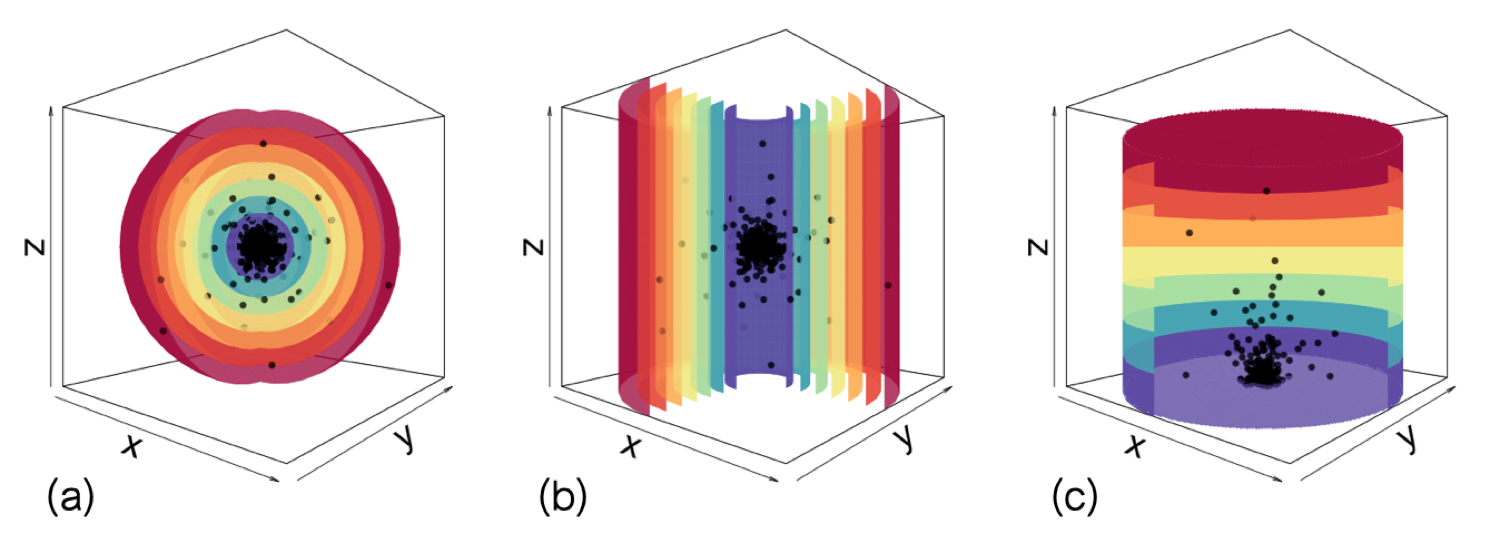}
\caption{(a) \texttt{bin\_type = ``r''} Demonstrating the 3D spherical bins. (b) \texttt{bin\_type = ``cr''} The  2D circular annuli bins that spread out radially along the plane of the disk. (c) \texttt{bin\_type = ``z''} The 2D circular bins that grow in 1D out of the plane of the disk.}
\label{fig:bin_dir}
\end{figure}

Particles are divided into bins (spherical shells, cylindrical shells or stacks - as shown in Figure \ref{fig:bin_dir}) and the following properties are computed:
\begin{itemize}
\item the mass distribution,
\item the log(density) distribution,
\item the circular and rotational velocity distributions,
\item the velocity anisotropy ($\beta$) distribution \citep{Binney2008GalacticDynamics},
\item the Bullock spin parameter ($\lambda$) distribution, \citep{Bullock2001AHalos}.
\end{itemize} 

These reflect the true nature of the system and allow us to explore the limitations of the synthetic observables - for example, when seeing conditions become more severe. 

\subsection{Creating the ``observable'' format}
\label{sec:method2}

In order to generate a projected galaxy image, as if the simulation is being observed in the sky, a few additional properties are added to each particle, such as the projected quantities of position and line-of-sight velocity at inclination, $i$, to the observer. 

\begin{align}
z_{obs} &= z \text{sin}(i) + y \text{cos}(i), \\
v_{los} &= v_z \text{cos}(i) - v_y \text{sin}(i), \\
r_{obs} &= \sqrt{x^2 + z_{obs}^2},
\end{align}

\noindent where $i = 0^\circ$ is the galaxy projected face-on and $90^\circ$ is edge on. 

\simspin{} then accounts for the physical properties of the observing telescope. Particulars such as the size and shape of the field of view, and the size of the galaxy within that aperture are specified. Further instrument specifics, such as CCD noise and detailed fibre arrangements, are not included in the current implementation, but we intend to add these in later iterations of the code. Using the \texttt{celestial} package\footnote{\url{https://CRAN.R-project.org/package=celestial}}, we compute the angular diameter size, $d_A$, of the galaxy when projected at a supplied redshift distance using equation \ref{eq:cosdis}. The reference cosmology in this case is the most recent Planck data \citep[H$_0 = 68.4$, $\Omega_M = 0.301$, $\Omega_L = 0.699$, $\Omega_R =  8.98 \times 10^{-5}$, $\sigma_8 = 0.793$;][]{PlanckCollaboration2018PlanckParameters}. 

\begin{equation}
\label{eq:cosdis}
d_A = \frac{S_k(r)}{1+z},
\end{equation}

\noindent where $r$ is the comoving distance and,

\begin{equation*}
S_k(r) = 
    \begin{cases}
      \frac{\text{sin}(\sqrt{-\Omega_k} H_0 r)}{H_0 \left|\Omega_k\right|}, & \text{if}\ \Omega_k < 0 \\
      r, & \text{if}\ \Omega_k = 0 \\
      \frac{\text{sin}(\sqrt{\Omega_k} H_0 r)}{H_0 \left|\Omega_k\right|}, & \text{if}\ \Omega_k > 0 \\
    \end{cases}
\end{equation*}

\noindent where $\Omega_k = 1-\Omega_M-\Omega_L-\Omega_R$ is the curvature density and H$_0$ is the Hubble parameter today.

Using this, we determine how large the galaxy appears within the aperture. A selection of aperture shapes are available (circular, hexagonal or square) to mimic the current layouts of modern IFS surveys, such as SAMI and MaNGA. The size of these apertures is user defined, specified by the diameter in units of arc-seconds. We remove any particles belonging to the galaxy that fall outside of the imaged region. 

Each remaining particle is then assigned a luminosity, $L$. This can be done by specifying a mass-to-light ratio for each luminous particle type (bulge, disc or star) and scaling the luminosity to a flux, $F$, with respect to the luminosity distance at a given redshift, $D_L = \sqrt{L / 4 \pi F}$; alternatively, if the user has computed a spectrum for each stellar particle, these can also be supplied to the code and used to calculate more accurate fluxes within a chosen filter using \prospect{}\footnote{\url{https://github.com/asgr/ProSpect}} \citep{Robotham2020ProSpect:Histories}, a high-level spectral generation package designed to create spectral energy distributions (SEDs) for the semi-analytic code, \textsc{SHARK} \citep{Lagos2018Shark:Formation}. \prospect{} combines stellar synthesis libraries, such as \cite{Bruzual2003Stellar2003} (BC03 hereafter) and/or EMILES \citep{Vazdekis2016UV-extendedGalaxies} with dust attenuation \citep{Charlot2000AGalaxies} and re-emission models \citep{Dale2014ANuclei}. In this case we use \prospect{} in a purely generative mode, using the SED generated for each stellar particle to calculate the flux contribution of each within a given filter.

The dimensions of the data cube are constructed to contain just the remaining luminous particles. We leave the specifics to the discretion of the user; for example, the apparent pixel size for SAMI data cubes is 0.5 arcsec with a spectral sampling scale of 1.04 \r{A} \citep{Green2018TheProducts}. These parameters are used to determine the widths of the bins in each direction. Physical pixel size is computed by multiplying the spatial sampling scale by the angular diameter scale calculated above, as this is dependent on the distance at which the galaxy is projected; the velocity pixel size is approximated by $\nu \sim c \Delta \lambda / \lambda$, where we take the central wavelength of the filter to be $\lambda$ and the spectral scale as $\Delta\lambda$. Particles are filtered into their correct positions within the position-velocity cube and output as a 3D array. 

At this stage, we make a key assumption: that each particle in the simulation has some inherent uncertainty in its velocity. This mimics the idea that, when astronomers observe emission lines, those lines have a width representing an uncertainty in the true speed of the host environment where the line originated. This uncertainty is encapsulated numerically by the line spread function (LSF), \texttt{lsf\_fwhm}, which is caused by a spectral response of the observing telescope to a point like source. We use the LSF to fix the ``width'' of the particle's velocity. As the LSF of IFS instruments can be well approximated as Gaussian, we model the velocity of each particle as a Gaussian centred on the known velocity of the simulated particle with a width corresponding to the LSF associated to the mock observation telescope \citep{vandeSande2017TheSurveys}.

\begin{figure*}[!ht]
\centering
\includegraphics[width=0.9\linewidth]{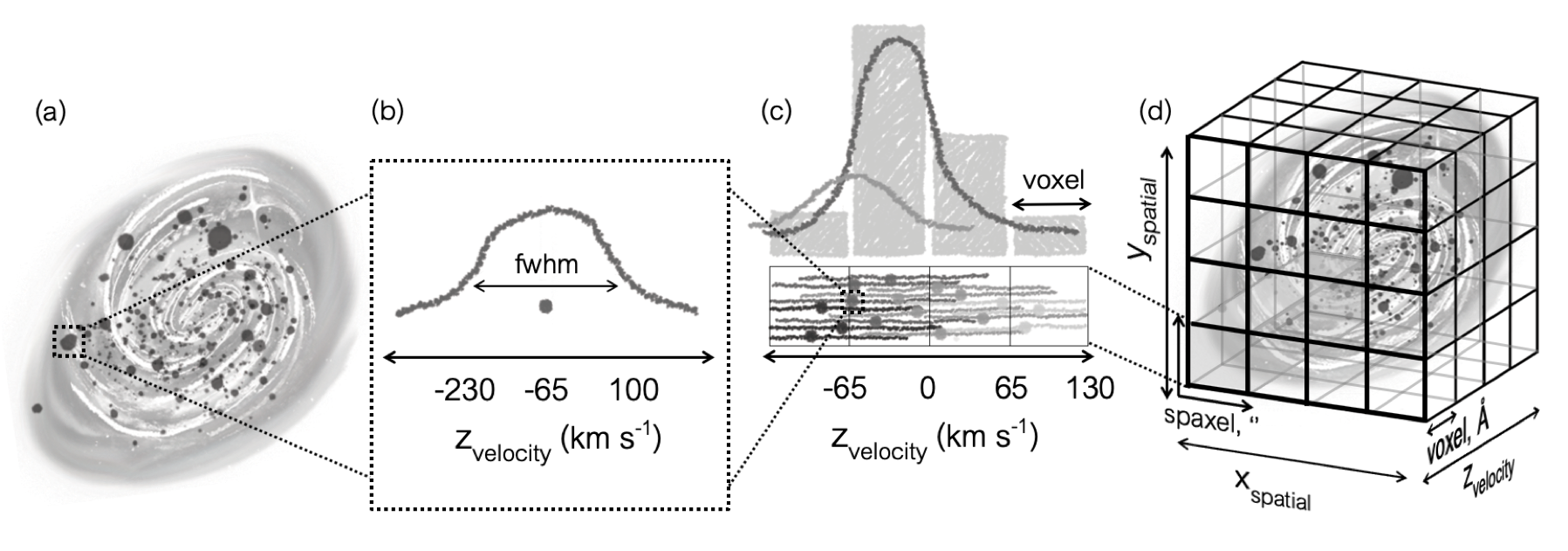}
\caption{Illustrating the method in which each kinematic data cube is constructed. (a) We take each particle within the simulation - which will have some known velocity along the projected LOS - and (b) convolve each with a Gaussian kernel such that it has a velocity distribution with width dictated by the LSF. (c) This velocity distribution is then binned in velocity space along each spatial pixel such that a single particle can occupy several velocity bins. (d) Each pixel is then arranged in the cube to reconstruct the galaxy image.}
\label{fig:build_datacube}
\end{figure*}

Each particle's associated Gaussian is scaled by the flux of that particle and then summed, with portions of each distribution contributing to several bins in velocity space. We fully bin the particles in this manner within both projected spatial coordinates and velocity space to construct the IFS kinematic data cube. An info-graphic of this process is shown in Figure \ref{fig:build_datacube}.

These arrays can be output in FITS file format, or passed to further functions for the addition of atmospheric effects and kinematic analysis. 

\subsection{Mimicking the effects of the atmosphere}
\label{sec:method3}

Ground-based optical observations are limited by the blurring effects of our atmosphere. In order to compare like-for-like, we replicate these ``beam smearing" effects within our synthetic observations. \simspin{} does this by convolving each spatial plane within the data cube with a point spread function (PSF).

The user can specify the shape of this PSF - either a Gaussian or Moffat kernel \citep{Moffat1969ANASA/ADS} - and the full-width half-maximum (FWHM) of the kernel. Each x-y spatial plane is convolved with the generated PSF, using functions from \textsc{ProFit} \citep{Robotham2017ProFitImages} which follow the method:
\begin{equation}
    F_{obs} = F_i \circledast \text{PSF},
\end{equation}
where $F_i$ is the flux within each pixel in each spatial plane, $i$, and $\circledast$ represents convolution. 

\subsection{Constructing synthetic images}
\label{sec:method4}
Having generated a realistic kinematic data cube, \simspin{} can be used to process images and observable kinematics. Flux images and maps of the LOS velocity and velocity dispersion are generated by collapsing the cube along the z-axis. 

\begin{figure}[!ht]
\centering
\includegraphics[width=\columnwidth]{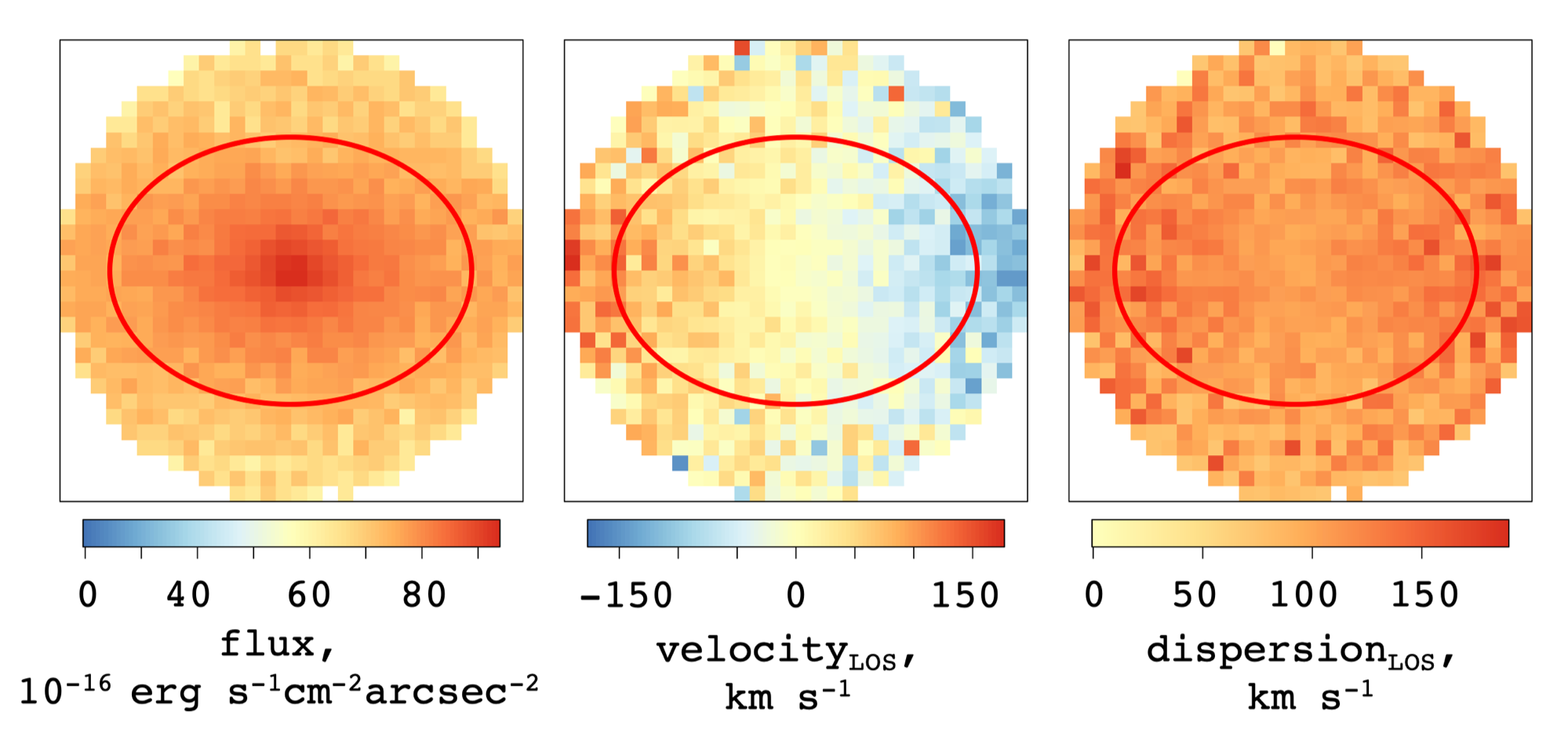}
\caption{Demonstrating the mock images produced through \texttt{SimSpin} observations of the S0 example model inclined to 70$^o$ with added Sky RMS noise. The red line demonstrates 1 R$_{eff}$, within which $\lambda_R$ is measured.}
\label{fig:find_lambda}
\end{figure}

To generate the flux maps, the contribution of flux from each velocity plane is summed, $F_i$:
\begin{equation}
  \label{eq:flux}
  F = \sum_{i = 1}^{v_{\text{max}}}{F_i},
\end{equation}
where $v_{\text{max}}$ is the last velocity bin along that pixel in the cube. The LOS velocity and LOS velocity dispersion are given by flux-weighted statistics:
\begin{align}
  \label{eq:LOS}
  V &= \frac{\sum{v_i \times F_i}}{\sum{F_i}}, \\
  \label{eq:LOSVD}
  \sigma &= \frac{\sum{F_i \times (v_i - V)^2}}{\sum{F_i}}.
\end{align}
Here, $v_i$ is the velocity assigned to each velocity bin, $i$, weighted by the flux in each bin, $F_i$, and $V$ is the mean velocity along that pixel in the cube given by eq \ref{eq:LOS}. An example of such images can be seen in Figure \ref{fig:find_lambda}.

Sky noise can optionally be added to the images by using a sample of random, Normally distributed values. The appropriate level is determined using the specified magnitude threshold, zero point and spatial pixel scale. It is possible to export these images for Voronoi binning. While this is not supported at this time within the \simspin{} code, \cite{Cappellari2003AdaptiveTessellations} provides a standard method for binning these images so that the signal-to-noise is consistent across pixels. \textsc{Vorbin}\footnote{\url{http://www-astro.physics.ox.ac.uk/~mxc/software/}} is a Python code that can be downloaded directly from PyPi. In future versions of this code, we intend for re-binned images to be added back in to \simspin{} for calculating the observable properties. 

\subsection{Calculating observable properties}
\label{sec:method5}
The synthetic images can then be used to calculate various observational kinematic properties of the galaxy in question. \simspin{} has been used to investigate the observable spin parameter, $\lambda_R$ \citep{Emsellem2007TheGalaxies,Harborne2019A_R}, and can further evaluate the $V/\sigma$ parameter \citep{Cappellari2007TheKinematics}.

The user can specify the radius within which the kinematic measurements are made. Often these are made within an effective radius, R$_{\text{eff}}$, but we give the user the freedom to fully specify the size and ellipticity of this measurement radius. Either, the second-order moments are calculated from the flux distribution by diagonalizing the inertia tensor and assuming the galaxy ellipticity from this axial ratio, $q$. This ellipse is then grown from the centre until half the total flux is contained within the radius. Alternatively, software like \textsc{ProFound}\footnote{\url{https://github.com/asgr/ProFit}} can be used to generate concentric isophotes that contain equal amounts of flux within each \citep{Robotham2017ProFitImages}. This axial ratio information can be used to specify the ellipse within which the kinematics will be calculated. Only the pixels whose midpoints are contained within this ellipse will be used for further calculations.

Currently, it is possible to calculate two kinematic properties: $\lambda_R$ \citep{Emsellem2007TheGalaxies} and V/$\sigma$ \citep{Cappellari2007TheKinematics}. $\lambda_R$ is calculated using Eq \ref{eq:lambdaR}:
\begin{equation}
\label{eq:lambdaR}
\lambda_R = \frac{\sum_{i=1}^{n_p} F_i R_i |V_i|}{\sum_{i=1}^{n_p} F_i R_i \sqrt{V_i^2 + \sigma_i^2}},
\end{equation}
where $F_i$ is the observed ``flux'' taken from the flux image, $R_i$ is the circularised radial position, $V_i$ is the LOS velocity taken from the LOS velocity image, $\sigma_i$ is the LOS velocity dispersion taken from the LOS dispersion image per pixel, $i$, and summed across the total number of pixels, $n_p$. We also give the option to compute the parameter using the elliptical radius where we define $R^{\epsilon}_i$ as the semi-major axis of an ellipse that would pass through that pixel.

Similarly, $V/\sigma$ is calculated:
\begin{equation}
\label{eq:vsigma}
    V/\sigma = \sqrt{\frac{\sum_{i=1}^{n_p} F_i V_i^2}{\sum_{i=1}^{n_p} F_i \sigma_i^2}}.
\end{equation}

\simspin{} computes these parameters in a consistent manner to observable software, such as pPXF \citep{Cappellari2017ImprovingFunctions}, for simple and consistent comparison with real observations. 

\section{IMPLEMENTATION}
\label{sec:implementation}

\begin{figure*}[!ht]
\centering
\includegraphics[width=0.85\linewidth]{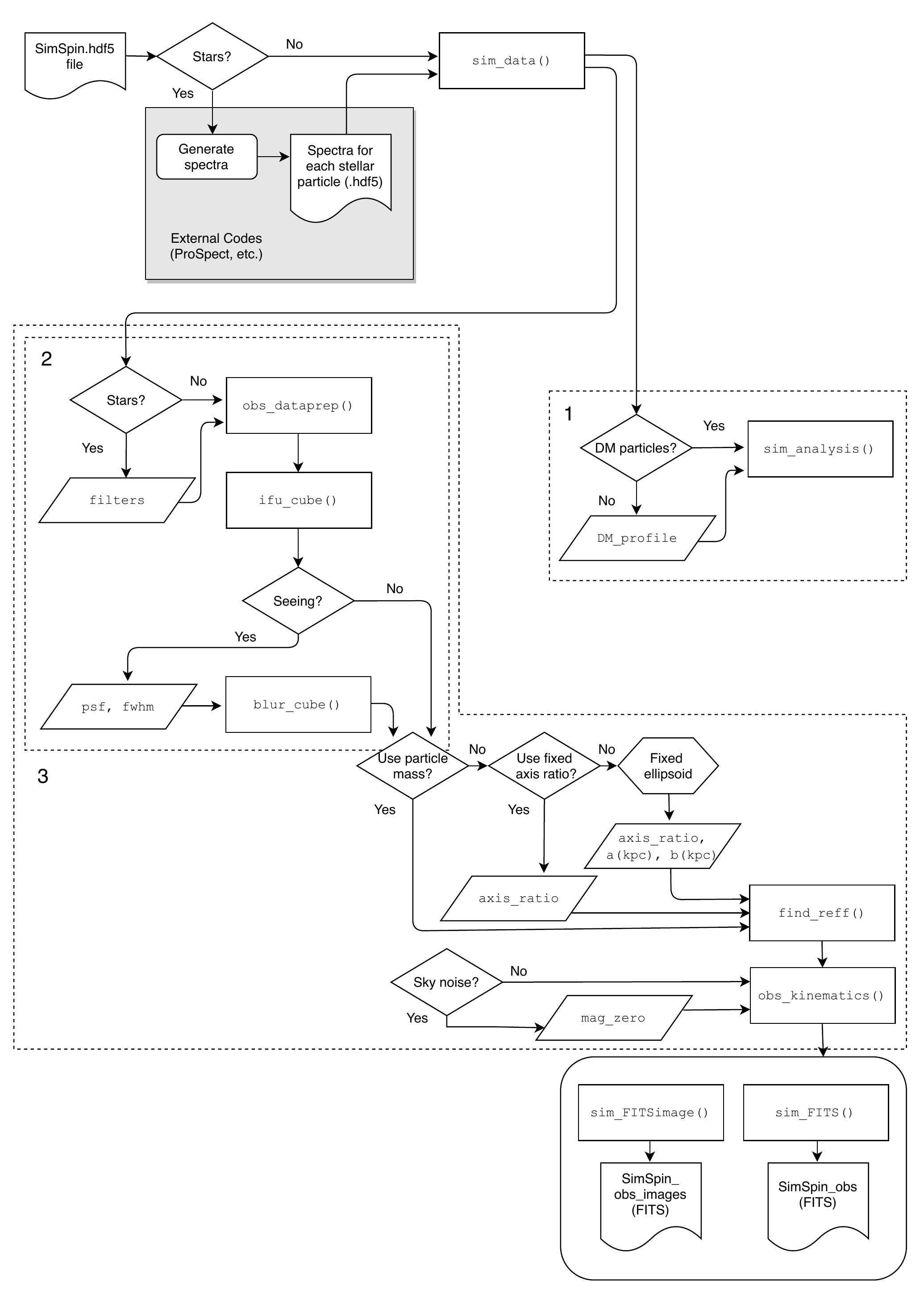}
\caption{Demonstrating the individual functions and queries used when running a simulated galaxy through the \simspin{} package. Three over-arching functions are identified that link these sub-functions together: (1) \texttt{sim\_analysis()} - as explained in Section \ref{sec:method1}, (2) \texttt{build\_datacube()} - as explained in Section \ref{sec:method2} and (3) \texttt{find\_kinematics()} - as explained in Section \ref{sec:method5}}
\label{fig:flow}
\end{figure*}

Having outlined the methodology, we present the fully documented and tested R-package, \simspin{}. This code can be downloaded from the {\small github} repository\footnote{\url{https://github.com/kateharborne/SimSpin}} and the package can be installed directly into R using the following commands:

\smallskip
\noindent \texttt{> install.packages("devtools") \\
> library(devtools) \\
> install\_github("kateharborne/SimSpin")}
\bigskip

\noindent To load the package into your R session, 

\smallskip
\noindent \texttt{> library(SimSpin)}
\bigskip

In Figure \ref{fig:flow}, we show the designed flow of the code. While it is possible to use each sub-function within this package independently and examine the output at each stage, there are three basic analysis functions designed to give the output information in a user friendly format.

\begin{enumerate}
\item \texttt{sim\_analysis()} - This function outputs the inherent kinematic properties of the galaxy model. This provides the comparison to the kinematic observables produced in the following functions. 
\item \texttt{build\_datacube()} - This function produces the kinematic data cube prior to kinematic analysis. This allows the user to take the cubes to use in some other form of analysis without having to calculate $\lambda_R$.
\item \texttt{find\_kinematics()} - This function produces a kinematic data cube and calculates the observed spin parameter, with both circularised and elliptical radii, V/$\sigma$, ellipticity, inclination and the corresponding flux, line-of-sight velocity and line-of-sight velocity dispersion images. For individual $\lambda_R$ or V/$\sigma$, we provide two functions (\texttt{find\_lambda()}/ \texttt{find\_vsigma()}) that output their named kinematics.
\end{enumerate}

For further information about the implementation of these functions, we direct the reader to the repository and to {\small RPubs} where we present a series of vignette examples\footnote{\url{https://rpubs.com/kateharborne}}. Each function is fully documented with an demonstrated example. 

\section{Examples}
\label{sec:examples}

In this section, we demonstrate two ways in which this package may be used. Section \ref{sec:ex1} shows the simple analysis of an $N$-body model containing disc and bulge components, and examines the effect of particle resolution on the profiles recovered. Section \ref{sec:ex2} expands further on this exercise to examine a hydrodynamic \eagle{} simulation with stellar particles and histories.

\subsection{\texorpdfstring{$N$}{N}-body model example}
\label{sec:ex1}

Here we use \simspin{} to measure the inherent kinematics and the observable spin parameter for five different N-body realisations of galaxies. We use repeated observations to explore how the particle resolution of an N-body model can impact the precision and reliability of the physical and observable kinematics recovered. 

These models have been constructed in two phases: first, initial conditions are generated using \galic{} \citep{Yurin2014AnEquilibrium} which constructs isolated particle distributions in equilibrium by solving the collision-less Boltzmann equation; we then evolve these initial conditions using a modified version of \gadget{} \citep{Springel2005TheGadget-2} in which the ``live'' dark matter (DM) halo is replaced with its ``static'' analytical form. This is done to maintain a stable, well-resolved galactic disk at reasonable computational cost. The details of this simulated catalogue can be found in Table \ref{tab:catalogue} and is described in greater detail within \citealt{Harborne2019A_R}.

\begin{table}[!ht]
\centering
\caption{Outlining the properties of each N-body galaxy model in the catalogue explored throughout the examples in Section \ref{sec:ex1}.}
\label{tab:catalogue}
\begin{tabular}{@{}lccccc@{}}
\toprule
 & B/T & b/kpc & n & N$_{disc}$ & N$_{bulge}$ \\ \midrule
\textbf{S0} & 0.60 & 2.14 & 2.84 & 1,000,000 & 1,500,000 \\
\textbf{Sa} & 0.40 & 1.38 & 2.26 & 1,500,000 & 1,000,000 \\
\textbf{Sb} & 0.25 & 0.90 & 1.64 & 1,875,000 & 625,000 \\
\textbf{Sc} & 0.05 & 0.17 & 0.99 & 2,375,000 & 125,000 \\
\textbf{Sd} & 0.02 & 0.07 & 0.97 & 2,450,000 & 50.000 \\ \bottomrule
\end{tabular}
\end{table}

First, we convert the simulation snapshots into \simspin{} compatible HDF5 input files\footnote{\url{https://github.com/kateharborne/create_SimSpinFile}} which can be read into R using the \simspin{} function \texttt{sim\_data()}, and the output data frame is processed using \texttt{sim\_analysis()} and \texttt{find\_lambda()}.  In each case, we examine our models at full particle resolution initially and assume this to be an ideal case by which we benchmark.

\subsubsection{Simulation properties}
We begin by examining the kinematic properties inherent to the simulated models. This is done using the code below. First we load in the simulation data, considering all components and then the disc and bulge components separately.
\bigskip

\noindent \texttt{> all\_data = sim\_data("S0.hdf5")}
\smallskip

\noindent \texttt{> disc\_data = sim\_data("S0.hdf5", ptype=2)} 
\smallskip

\noindent \texttt{> bulge\_data = sim\_data("S0.hdf5", ptype=3)} 
\bigskip

Next, we run the \texttt{sim\_analysis()} function for each loaded data set, supplying the information about the DM profile that has been removed throughout the evolution and the number of radial bins that we wish to examine the profile across. If DM particles were present in the supplied file, the DM profile parameter would not be necessary. 
\bigskip

\noindent \texttt{> all\_analysis = sim\_analysis(all\_data, rbin = 1000, DM\_profile = list(profile=``Hernquist'', DM\_mass=184.996, DM\_a=34.5))}
\medskip

\noindent \texttt{> disc\_analysis = sim\_analysis(disc\_data, rbin = 1000, DM\_profile = list(profile=``Hernquist'', DM\_mass=184.996, DM\_a=34.5))}
\smallskip

\noindent \texttt{> bulge\_analysis = sim\_analysis(bulge\_data, rbin = 1000, DM\_profile = list(profile=``Hernquist'', DM\_mass=184.996, DM\_a=34.5))} 

\bigskip

\begin{figure}[!ht]
\centering
\includegraphics[width=\columnwidth]{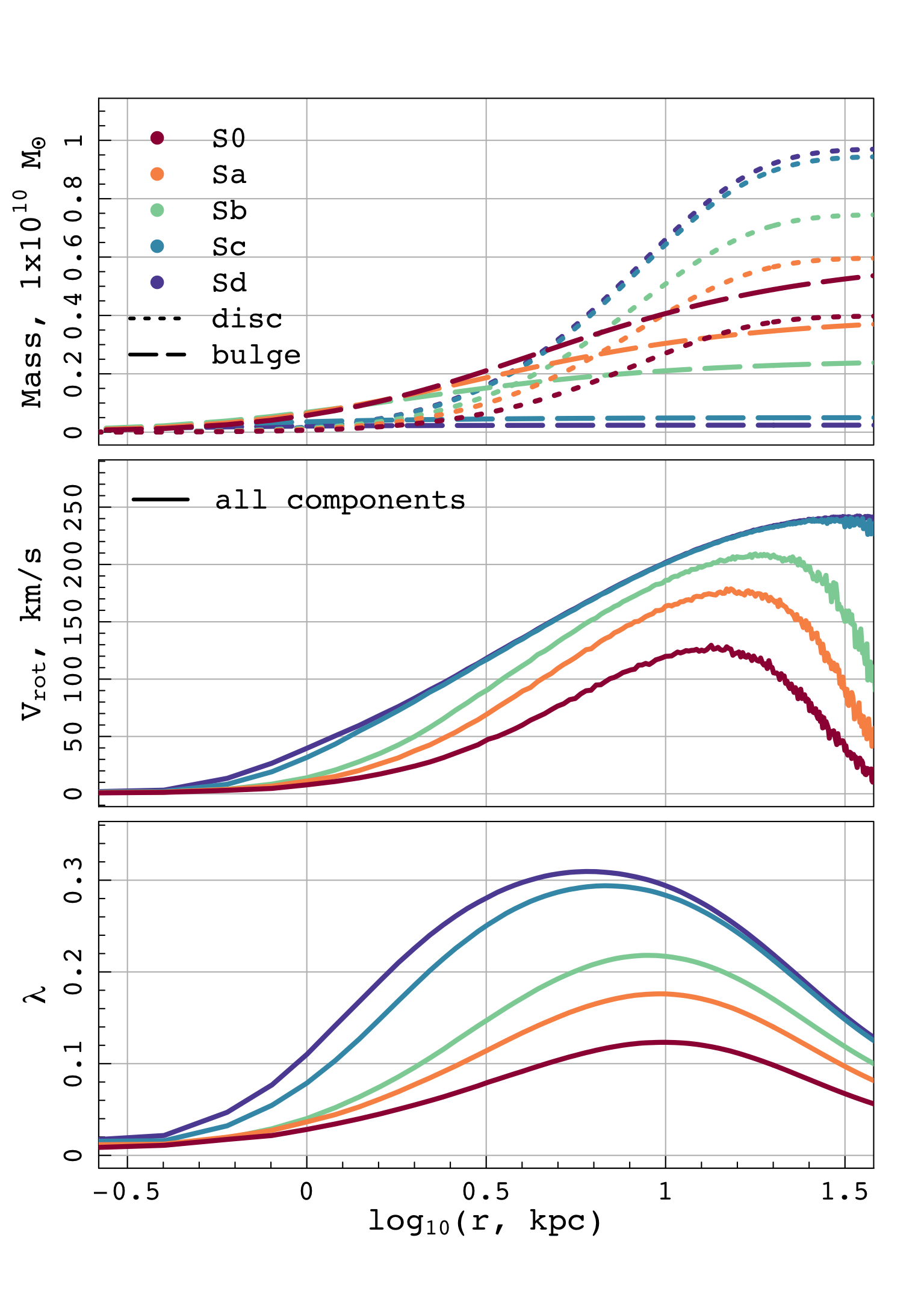}
\caption{Showing a selection of the profiles provided by the \texttt{sim\_analysis()} function: Mass (top), rotational velocity (middle), Bullock spin parameter (bottom). We demonstrate the flexibility of the code in analyzing galaxy components separately.}
\label{fig:sim_analysis}
\end{figure}

There are several outputs from the \texttt{sim\_analysis()} function, as described in Section \ref{sec:method1}. Some of these are shown in Figure \ref{fig:sim_analysis}. We demonstrate how simply we can examine different simulation components in the top panel, where the mass profile of the disc and bulge are plotted separately for each galaxy in the catalogue. 

We have reduced the resolution of our simulations in order to examine what effect this has on the recovered kinematics. For each test, we have rerun the \galic{} initial conditions of the galaxy but with a fraction of the original number of particles. These have then been evolved for 10 Gyrs using \gadget{}. We have considered a further 12 iterations of each galaxy in the catalogue, from 0.01\% N$_{\text{total}}$ increasing in increments to the full N$_{\text{total}}$. This gives a sample of 65 simulations to analyse overall. 

\begin{figure}[!ht]
\centering
\includegraphics[width=\columnwidth]{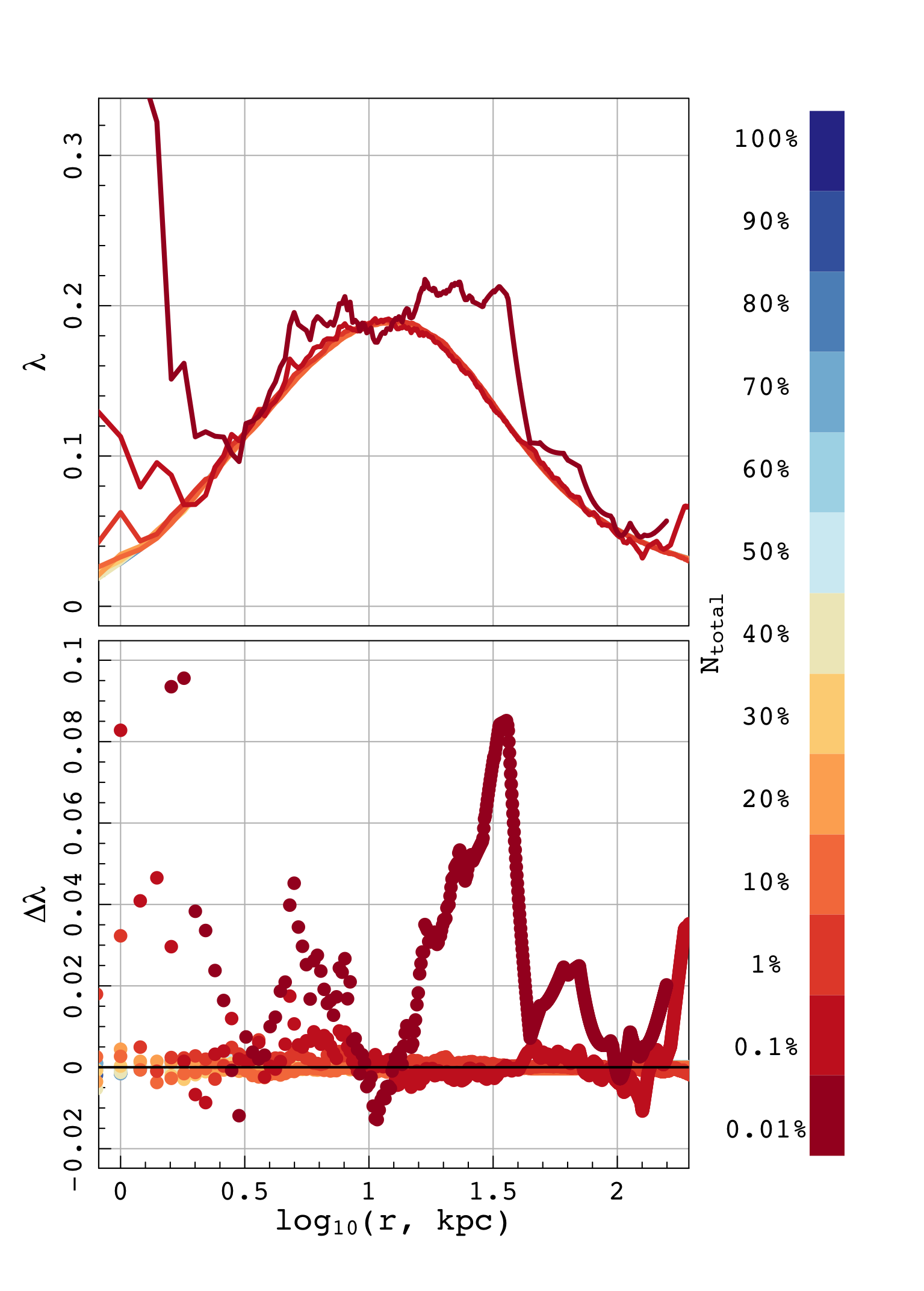}
\caption{Demonstrating the effect of reduced particle resolution on the recovery of the Bullock spin parameter, $\lambda$. In the upper panel, we show the spin parameter radial profile for the Sa galaxy at 13 different resolutions, as described by the colours shown on by the colour bar on the right. The percentages describe the fraction of the N$_{\text{total}}$ particles included in each simulation. Residuals from the 100\% N$_{\text{total}}$ case are shown in the lower panel.}
\label{fig:sim_analysis_res}
\end{figure}

The results of this experiment are shown for the measured Bullock spin parameter in Figure \ref{fig:sim_analysis_res} for the Sa galaxy. As we expect, the lower the resolution, the poorer our results become. Clearly, for the lowest resolution considered, at 250 particles, we have very noisy variations ($\pm 0.1$) between measured lambda and the benchmark measurement made at full resolution. However, these variations are much less significant at 10\% N$_{\text{total}}$, where we only see $\pm0.005$ and only within the inner radii. In all but the worst case, the expected shape of the lambda profile is still recovered. The effect of particle resolution is akin to adding numerical noise. While the general trend is returned, we see the effective signal-to-noise drop as the number of particles used to describe the mass distribution is reduced. It is interesting to investigate how the observable kinematics are affected by similar reductions in resolution. 

\subsubsection{Synthetic observable properties}

Using the \texttt{find\_lambda()} function, we can generate synthetic observations of our models and measure kinematic properties. The following code generates images as if our simulations were observed using the blue arm of SAMI, which is the default output of the \simspin{} observation functions. The projected distance, $z$, is set to 0.05 and the projected inclination for these observations is $70^{\circ}$.

\bigskip

\noindent \texttt{> all\_data = sim\_data("S0.hdf5")}
\smallskip

\noindent \texttt{> lambda = find\_lambda(all\_data)}

\bigskip

\begin{figure}[!ht]
\centering
\includegraphics[width=\columnwidth]{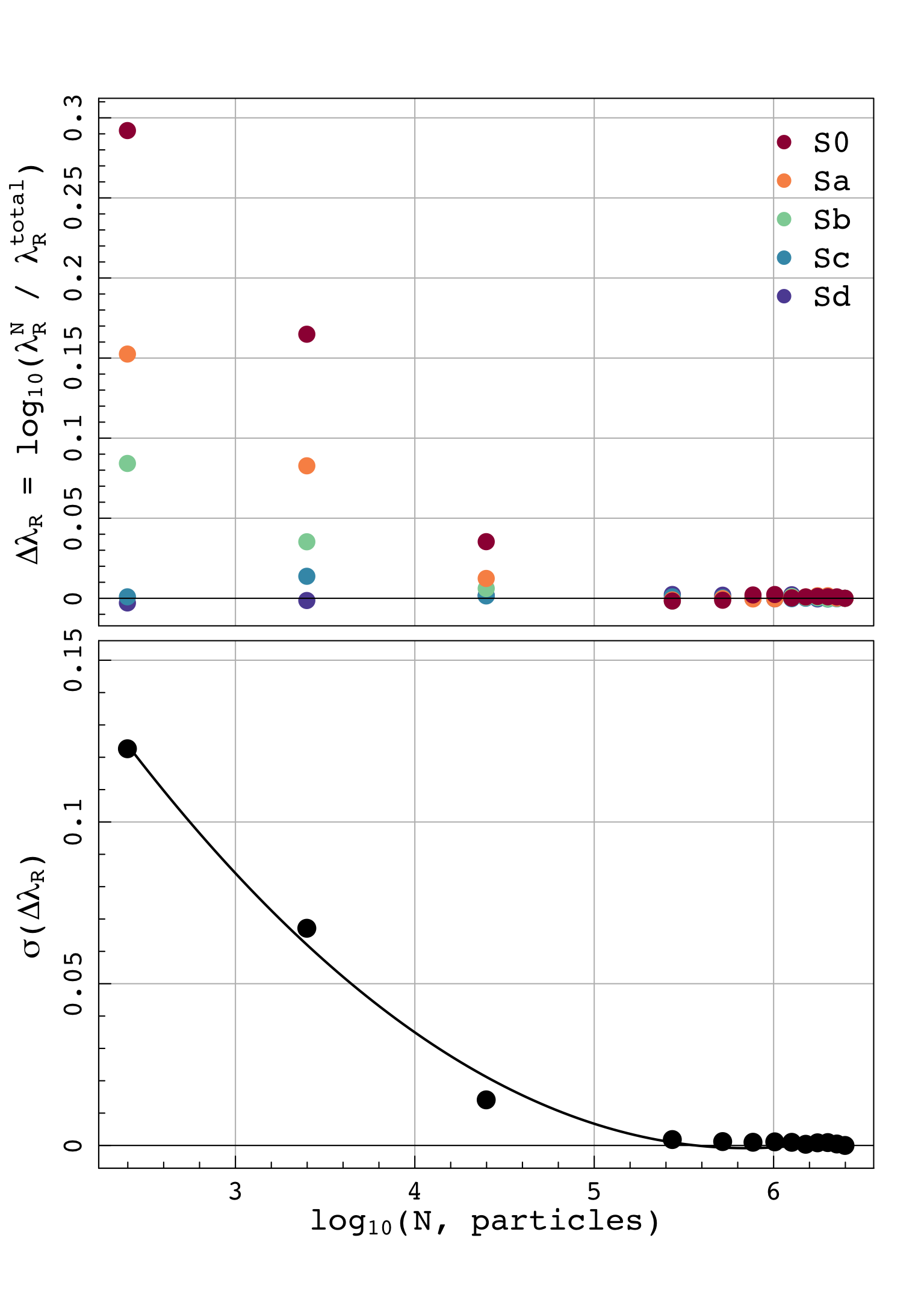}
\caption{Considering how the observed spin parameter, $\lambda_R$, changes when the number of particles within the simulated model is reduced from full resolution down to 0.01\% N$_{total}$. In the upper panel, we consider the log difference between $\lambda_R$ at that resolution with respect to the value measured at the benchmark resolution. In the lower panel, we measure the scatter of those measurements at each resolution and find that it is well fit by an exponential profile.}
\label{fig:deltalr_Ntotal}
\end{figure}

We run this function on the 12 degraded iterations of each galaxy model in the catalogue, examining each spin parameter at resolutions from 0.01\% to 100\% of the benchmark. In each case, the recovered $\lambda_R$ is plotted against the particle resolution; this is shown in Figure \ref{fig:deltalr_Ntotal}. In the upper panel of the figure, we examine the log difference $\Delta\lambda_R = \text{log}_{10}(\lambda_R^{N} / \lambda_R^{total})$, i.e. between the value of $\lambda_R$ observed at a degraded resolution and the value measured at the highest resolution. In the lower panel of this figure, we consider the spread of the $\Delta\lambda_R$ values at each resolution. Within each bin we measure the standard deviation, $\sigma$, in order to give an indication of how confident we are that the model returns the correct $\lambda_R$. This assumes that the full resolution, $2.5 \times 10^{6}$-particle model is the ``true'' value. 

From the lower panel in Figure \ref{fig:deltalr_Ntotal}, we can see that at resolutions lower than $5 \times 10^5$ particles, the measure becomes exponentially more noisy. At 0.01\% N$_{total}$ particles, $\sigma$ peaks at $\sim 0.12$. However, the change in $\sigma(\Delta\lambda_R)$ is quite sudden, when the number of particles in the model approaches $5 \times 10^5$. At higher resolutions than this, we see that the benchmark $\lambda_R$ is recovered with a precision of $\sim0.002$, which is much lower than normal observational uncertainties. The assumption that the $2.5 \times 10^{6}$-particle model returns a benchmark value also seems reasonable. The fairly flat trend in $\sigma(\Delta\lambda_R)$ from 30\% - 100\% N$_{total}$ suggests that we have reached an asymptotic value.

We also see that as the number of particles drops, the morphology of the model becomes important for the associated uncertainty. At 0.01\%, the S0 galaxy has the greatest uncertainty. This makes sense when you consider that dispersion will dominate in the bulge component; reducing the number of bulge particles that sample this velocity distribution will have a greater impact on the measurement of $\sigma$, and hence $\lambda_R$ and so we see the largest uncertainties in galaxies with larger bulge components.

Overall, this demonstrates that galaxies represented by smaller numbers of particles will give less accurate measurements, though in this example the uncertainty on the measurement at N > 250,000 is $\sim0.002$, increasing to $\sim0.02$ as we approach smaller N, $\sim 25,000$. In reality, even at this number of particles, the related uncertainty is still much smaller than the uncertainty generally associated with observational limitations (i.e. seeing and beam smearing) \citep{DEugenio2013FastZ=0.183, vandeSande2017TheSurveys, vandeSande2017TheKinematics, Greene2018SDSS-IVGalaxies, Graham2018SDSS-IVProperties, Harborne2019A_R}.

\subsection{Full hydrodynamical model example}
\label{sec:ex2}

It is important to understand how uncertainties are associated with particle resolution because when we examine galaxies from larger cosmological models, the number of particles making up individual galaxies tends to be lower. In the following example, we use a small galaxy from the \eagle{} simulation, \texttt{GalaxyID = 1056} taken from the \texttt{RefL0100N1504} simulation. As a proof of concept, we will demonstrate that we can process these galaxies using the same functions as the simple N-body models in Section \ref{sec:ex1}.

The \eagle{} project \citep{Schaye2015TheEnvironments, Crain2015TheVariations, McAlpine2016TheCatalogues} is a suite of cosmological hydrodynamical simulations that allows us to investigate the formation and evolution of galaxies. We have been using the publicly available \texttt{RefL0100N1504} simulation run; this is a cubic volume with a side length of 100 co-moving Mpc, of intermediate resolution with initial baryonic particle masses of $m_g = 1.81 \times 10^6$ M$_{\odot}$, and maximum gravitational softening lengths of $\epsilon_{\text{prop}} = 0.70$ pkpc. 

\begin{figure}[!ht]
\centering
\includegraphics[width=\columnwidth]{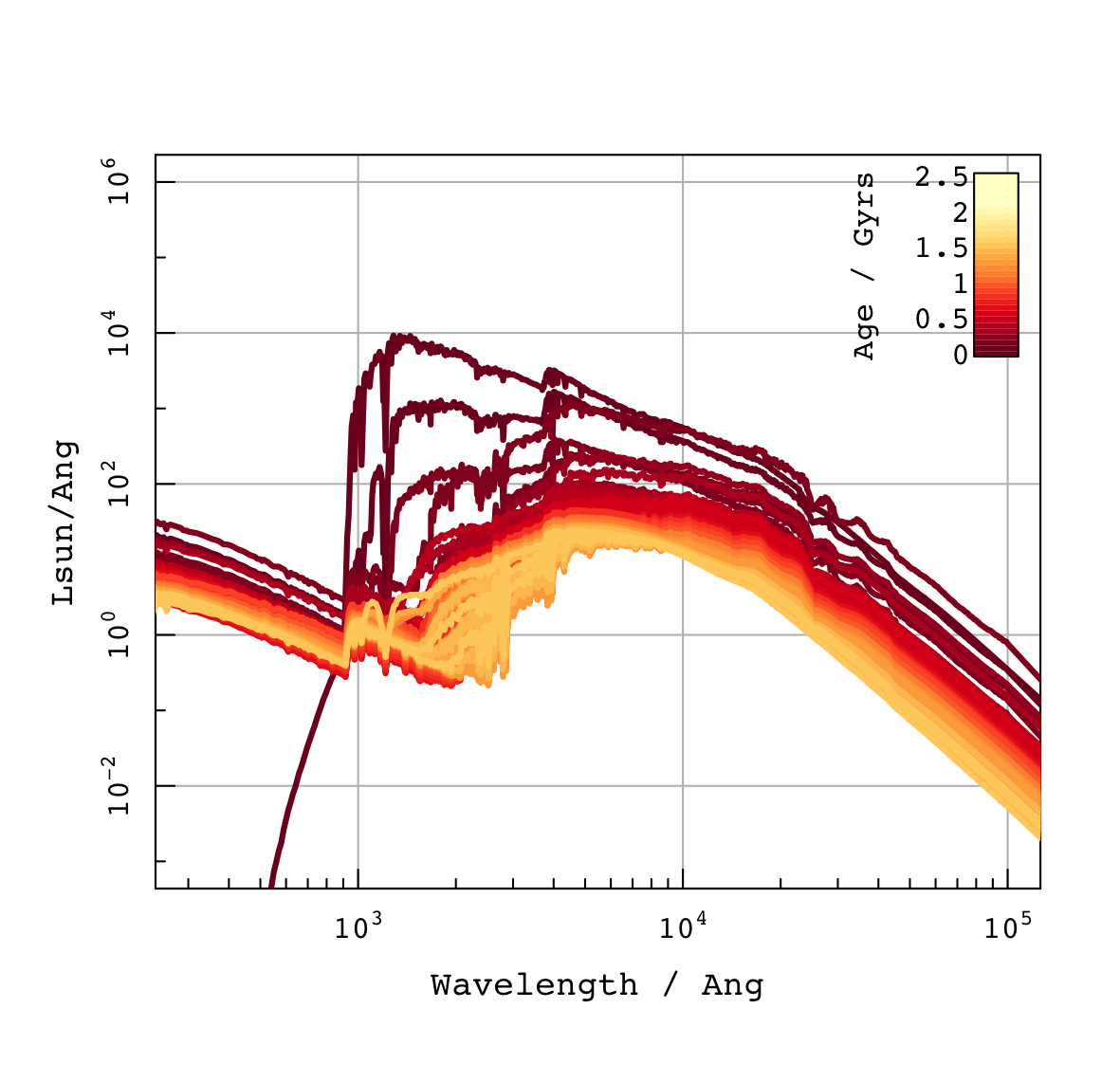}
\caption{Showing the SEDs generated using \prospect{} for 500 of the stellar particles within \eagle{} \texttt{GalaxyID = 1056}. The colour of each line reflects the age of the stellar particle in Gyr. These particles have been randomly selected from the 21,174 within the model to show a representative number of the stellar distributions in this galaxy.}
\label{fig:prospect}
\end{figure}

The galaxy chosen, \texttt{GalaxyID = 1056}, contains 21,174 stellar particles. In \eagle{}, each stellar particle is initialized with a stellar mass described by the \cite{Chabrier2003GalacticFunction} initial mass function; their metallicities are inherited from their parent gas particle and their ages recorded from their formation to current snapshot time. We can use this information to generate particle luminosities assuming a single stellar population (SSP). When providing \simspin{} with stellar particles, it is optional as to whether you would like to provide spectral energy distributions (SEDs) generated using stellar population synthesis models or just assume a mass-to-light ratio for all stellar particles. For this example, we have used \prospect{} to generate SEDs for each stellar particle by interpolating the BC03 tables in both age and metallicity. A subset of these generated distributions are shown in Figure \ref{fig:prospect}.

We store these SEDs in a \simspin{} compatible stellar file and provide this information along with the simulation data. This stellar file is in an HDF5 format that contains a \texttt{PartType4} group with two data-sets; \texttt{Luminosity} and \texttt{Wavelength}. When loading this data into \simspin{}, we specify this stellar file as below.

\bigskip

\noindent \texttt{> eagle\_data = sim\_data("eagle.hdf5", SSP="eagle\_stars.hdf5")}

\bigskip

This prepares a structure that contains the luminosity at a range of wavelengths for the stellar particles. We then pass this data structure on to the \texttt{find\_vsigma()} function as we would any other \texttt{sim\_data()} output.

\bigskip

\noindent \texttt{> eagle\_vsigma = find\_vsigma(eagle\_data, z=0.0005, inc\_deg=90)}

\bigskip

 The presence of these luminosity/wavelength tables means that, once passed to the \simspin{} analysis functions, the particle fluxes will be calculated using \prospect{} within a given filter. In this case, given that the central wavelength of the blue arm of SAMI ($\sim$4800\r{A}), we use the SDSS (Sloan Digital Sky Survey) g-filter to assign a flux to each particle at the specified redshift. Currently, this process is the most computationally expensive stage because the flux calculation takes on average $\sim0.004$s per particle. When increasing the number of particles above the small number in this simulation, this becomes the main source of computation, mostly due to the large arrays of luminosity and wavelength for each particle. For the purpose of proving that \simspin{} can perform with hydrodynamic simulations, we believe this to be sufficient, but highlight this as an area for further work that would benefit from parallelisation.

\begin{figure}[!ht]
\centering
\includegraphics[width=\columnwidth]{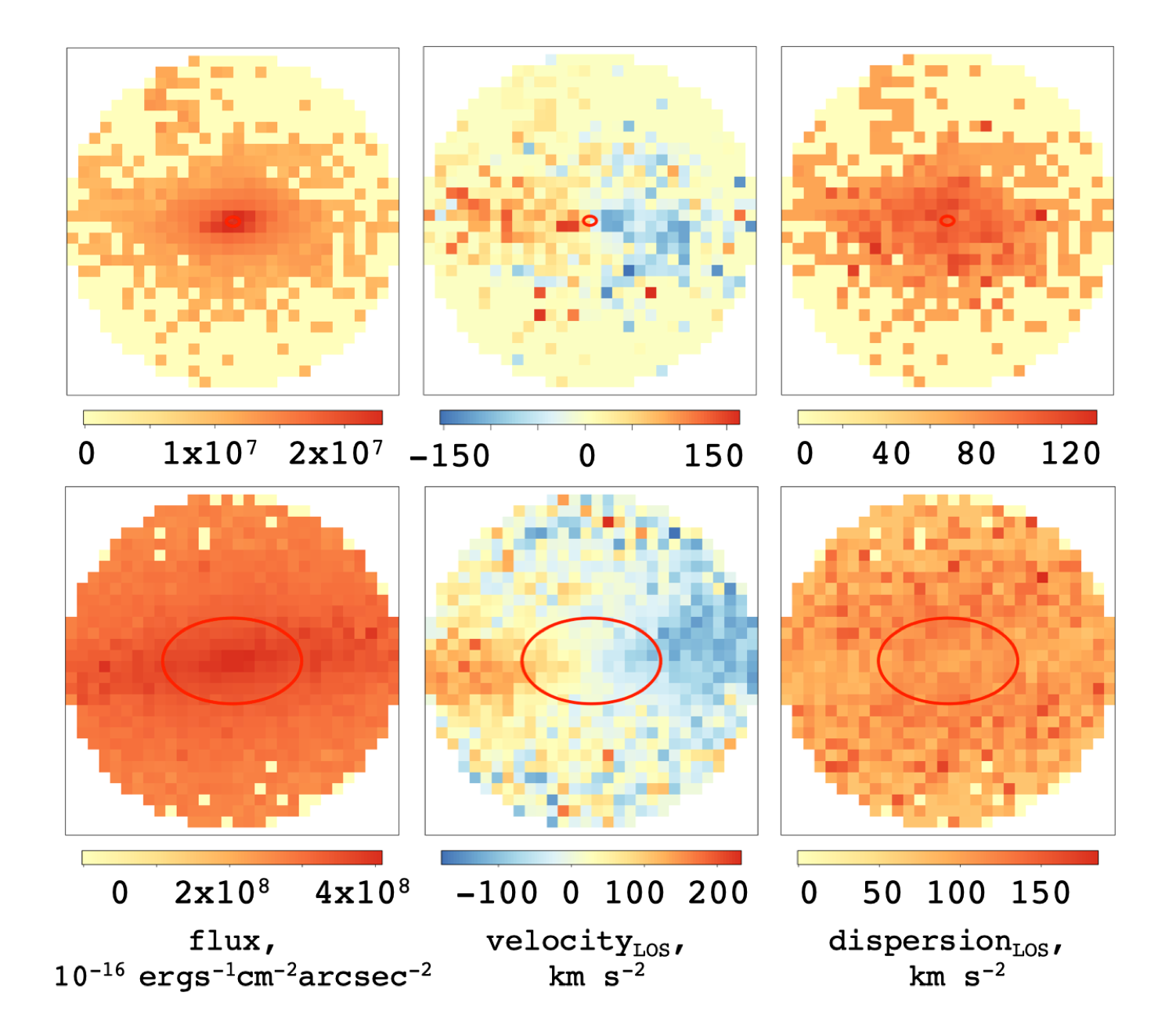}
\caption{The synthetic observations of the \eagle{} galaxy (\texttt{GalaxyID = 1056}) inclined edge-on, produced by \simspin{} at a projected redshift distance $z=0.0005$ (above) and $z=0.00005$ (below).}
\label{fig:EAGLE_observations}
\end{figure}

Because the galaxy in question is very small (R$_{\text{eff}} \sim 0.003$ kpc) we have had to put this galaxy very nearby in order to generate a suitable image. We show the images produced in Figure \ref{fig:EAGLE_observations}. Inclining the galaxy to 90$^{\circ}$ we see that there is a rotating disc component present. However, the image resolution is insufficient to make a measurement of $V/\sigma$ within a half-mass radius, as this makes up less than 4 pixels at the centre of the image. If instead we bring the galaxy closer again:

\bigskip

\noindent \texttt{> eagle\_vsigma\_close = find\_vsigma(eagle\_data, z=0.00005, inc\_deg=90)},

\bigskip
\noindent we now recover the $V/\sigma = 0.46$. Both projections are shown in Figure \ref{fig:EAGLE_observations}, with the fitted half-mass radius shown. 

\begin{figure*}[!ht]
\centering
\includegraphics[width=\linewidth]{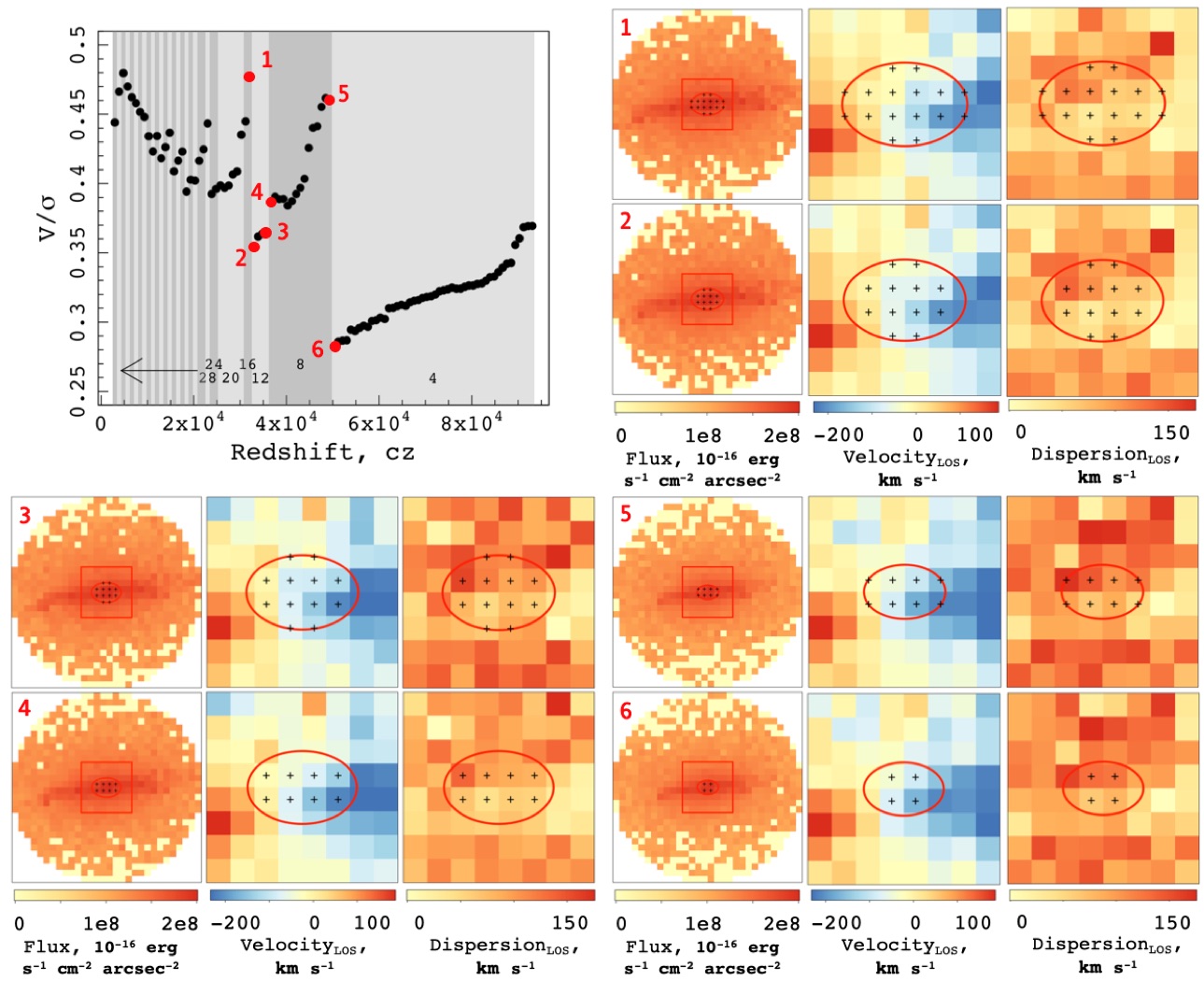}
\caption{Investigating the scatter in the measurement of $V/\sigma$ with 100 redshift distances from $z = 0.00001$ to $0.00031$ for the \eagle{} galaxy (\texttt{GalaxyID = 1056}).}
\label{fig:EAGLE_z}
\end{figure*}

This is another way in which resolution can effect our recovery of the internal kinematics. With instruments like {\small MUSE}, there are a large number of poorly resolved objects at distant redshifts for which we have a few pixels worth of kinematic information \citep{Guerou2017The0.8}. Using \simspin{}, we can examine what effect this has on the recovered kinematics. With this in mind, we varied the projected distance of the galaxy inclined edge-on from $z=0.00001$ to the point at which the measurement radius is too small to return a value (i.e. at the point when the semi-major axis of the measurement ellipse is less than the size of one pixel in the image). We present this in units of $cz$ for clarity (i.e. from $cz = 3,000 - 93,000$). This is shown in Figure \ref{fig:EAGLE_z}. 

There are several interesting features in this plot. Firstly, there are large discontinuities where the $V/\sigma$ measure drops suddenly. Within each of the discrete chunks, there is a gradual positive inclination of points. To explain these features, we have looked more closely at the geometry of the images being produced for each measurement.

Throughout this experiment, we have fixed the shape of the measurement ellipse in physical space ($a = 0.003$ kpc and $b = 0.002$ kpc) as was measured using the second-order moments of the galaxy projected at 90$^{\circ}$. Therefore, the only thing changing as we modify the projected redshift distance is the size of the galaxy within the aperture. As the galaxy is moved further and further from the telescope, a smaller number of pixels fit inside the measurement ellipse. Also, each pixel begins to contain more particles within that system, which explains the gradual increases we see within the discrete bins; to explain the discrete bins themselves, we first considered the number of pixels that are contained within the ellipse in each case. As can be seen by the shaded blocks behind the measurements, the jumps occur at the transitions when the ellipse crosses the midpoints of new pixels to be included in the measurement.

Finally, we see that the jumps can be positive or negative, and that in general these jumps become larger the further away we go. To understand this effect, we looked at the arrangement of additional particles in each of three cases. In the first, we consider the transition from point 1 to 2, as shown in Figure \ref{fig:EAGLE_z}. Here we see $V/\sigma$ drop considerably from 0.48 to 0.35; the corresponding maps produced show that at this point, we go from 16 to 12 pixels, and specifically we loose pixels along the semi-major axis. We have aligned the major velocity axis horizontally so the removal of the pixels in these positions reduces the velocity component much more than the dispersion component. Therefore, in a relative sense, the dispersion becomes larger and $V/\sigma$ drops. Conversely, if we remove pixels from the vertical semi-minor axis, we reduce the dispersion component relative to the velocity and so $V/\sigma$ rises; this is shown in the highlighted 3 to 4 points. The most significant drop is between points 5 and 6. At this stage we are hitting the limit of our resolution because we cannot calculate $V/\sigma$ with fewer than 4 pixels. This jump is the largest because more particles are contained within these 4 pixels than at the closer projections.

This result will be dependent on several other factors such as the inclination and morphology of the galaxy being observed. The magnitude of these jumps will also be changed by the manner in which you choose to centre your pixels (for example, if the centre of your measurement ellipse is at the centre of a single pixel), although we would still expect to see these jumps due to the discrete nature of pixel resolution. In all cases, this is pause for thought when calculating kinematic measures with only a very small number of pixels at the centre of your system.

\section{Further work \& conclusions}
\label{sec:further}

We have introduced \simspin{}, a flexible and versatile framework for generating and analysing IFS data cubes from $N$-body/hydrodynamical simulations, which can be compared directly with observations from surveys such as SAMI and MaNGA.

We offer this current version of \simspin{} via direct download from GitHub, or a web application\footnote{The app is housed on the Nimbus server at the Pawsey Supercomputing Centre, \url{http://simspin.icrar.org/}}. The benefit of using \texttt{R} is the ease with which an app can be created, and the \texttt{R Shiny} application makes it simple to generate a standard \texttt{JavaScript}, web browser-compatible GUI. The current version of this web app is designed for simplicity, and so there are fewer code options, described by a series of drop down panels and slider bars, than available via the standard \texttt{R} interface. The outputs generated are basic, but allow you to intuitively explore how observational effects such as seeing conditions and projected distance affect the observable kinematics.  This application can also be downloaded to your own system and run from there, available on {\small GitHub}\footnote{\url{https://github.com/kateharborne/SimSpin_app}}.

There are several aspects that we wish to develop further and in this section we will outline a few of these ideas. 

\medskip

\noindent In the near term, we will:

\begin{itemize}
\item increase the efficiency and reduce the run-time of the conversion between intrinsic spectrum and flux, which currently requires a large portion of computation time, as highlighted in the \eagle{} galaxy example in Section \ref{sec:ex2}.  
\item add further kinematic observables, such as higher order Gauss-Hermite coefficients, $h_3$ and $h_4$, and output images, such as gas maps as we continue to implement improvements for processing hydro-dynamical simulations. 
\end{itemize}

\medskip

\noindent In the longer term, we plan to:

\begin{itemize}

\item offer mock data cube outputs in a format that more closely mimics their observed counter parts, i.e. cubes with ($x,y,\lambda$) rather then the current ($x,y,v_{\text{LOS}}$). With the progression of tools such as \textsc{ProSpect}, each particle in a model can be associated with a spectra, as demonstrated in Section \ref{sec:ex2}. With these spectra used to describe the third dimension of the mock-cube, we can generate a product that can be fully processed using observational pipelines.

\item further signal-to-noise treatments could be implemented at the level of the individual spectra. Signal-to-noise values are important for observational kinematic measurements and the current implementation of \simspin{} does not yet incorporate this factor in a physically meaningful way.

\item additional instrument specifics and uncertainties can be implemented - CCD read-out noise, fibre arrangements, dithering patterns, etc.

\item provide multi-language implementations of \simspin{}, in, for example, \texttt{python} and \texttt{Julia}. Such versions are under development. Ideally, we would like several wrappers to allow different users to generate the same results easily from different platforms, such as a \texttt{python} wrapper using the \texttt{r2py} interface. 

\end{itemize}

Finally, one of the key requirements of this code has been that it should be easily extendable. All of the software is freely available through the \texttt{GitHub} repository and, with the use of this paper, examples within the package and \texttt{RPubs}, we hope that general users can extend this work to address many questions beyond those we have considered here. 

\medskip

In conclusion, we have demonstrated the versatility and simplicity of the \simspin{} code. While this tool was initially created to measure the observational effects that come into play when measuring kinematics such as $\lambda_R$, its extensibility makes the scope of this tool enormous. 
We have demonstrated examples in which we consider how the resolution of simulated models and the resolution of the observing instrument used may impact measurements of galactic kinematics. These two examples barely scratch the surface of all the parameter variations that could be considered.

We believe that \simspin{} has uses within both the theoretical and observational astronomical communities. The code has the ability to produce synthetic data products in FITS file format that allows direct comparison between simulations and observations. This code has already been used to examine how the observed kinematics vary with seeing conditions \citep{Harborne2019A_R}, and further projects include using synthetic \simspin{} observations to design an empirical correction to counteract these effects (Harborne et al., submitted). We present version 1.1.1 with the capabilities outlined within this paper, and invite further collaboration to extend the reach of this tool. 

\begin{acknowledgements}

We would like to thank the anonymous referee for their kind and prompt review of this work. KH is supported by the SIRF and UPA awarded by the University of Western Australia Scholarships Committee. CP and AR acknowledge the support of ARC Discovery Project grant DP140100395. Parts of this research were also conducted by the Australian Research Council Centre of Excellence for All Sky Astrophysics in 3 Dimensions (ASTRO 3D), through project number CE170100013. This work was supported by resources provided by the Pawsey Supercomputing Centre with funding from the Australian Government and the Government of Western Australia.

We acknowledge the Virgo Consortium for making their simulation data available. The \eagle{} simulations were performed using the DiRAC-2 facility at Durham, managed by the ICC, and the PRACE facility Curie based in France at TGCC, CEA, Bruy\`{e}resle-Ch\^{a}tel.

\end{acknowledgements}

\begin{appendix}

\end{appendix}

\bibliographystyle{pasa-mnras}
\bibliography{simspin.bib}

\begin{thebibliography}{}
\makeatletter
\relax
\def\mn@urlcharsother{\let\do\@makeother \do\$\do\&\do\#\do\^\do\_\do\%\do\~}
\definecolor{darkblue}{rgb}{0,0,0.597656}
\def\mndoi{\begingroup\mn@urlcharsother \@ifnextchar [ {\mndoi@} {\mndoi@[]}}
\def\mndoi@[#1]#2{\def\@tempa{#1}\ifx\@tempa\@empty \href
  {http://dx.doi.org/#2} {\textcolor{darkblue}{doi:#2}}\else \href
  {http://dx.doi.org/#2} {\textcolor{darkblue}{#1}}\fi \endgroup}
\def\mn@eprint#1#2{\mn@eprint@#1:#2::\@nil}
\def\mn@eprint@arXiv#1{\href {http://arxiv.org/abs/#1} {{\tt arXiv:#1}}}
\def\mn@eprint@dblp#1{\href {http://dblp.uni-trier.de/rec/bibtex/#1.xml}
  {dblp:#1}}
\def\mn@eprint@#1:#2:#3:#4\@nil{\def\@tempa {#1}\def\@tempb {#2}\def\@tempc
  {#3}\ifx \@tempc \@empty \let \@tempc \@tempb \let \@tempb \@tempa \fi \ifx
  \@tempb \@empty \def\@tempb {arXiv}\fi \@ifundefined
  {mn@eprint@\@tempb}{\@tempb:\@tempc}{\expandafter \expandafter \csname
  mn@eprint@\@tempb\endcsname \expandafter{\@tempc}}}

\bibitem[\protect\citeauthoryear{Ascasibar, Yepes, Gottl{\"{o}}ber  \&
  M{\"{u}}ller}{Ascasibar et~al.}{2002}]{Ascasibar2002NumericalHistory}
Ascasibar Y.,  Yepes G.,  Gottl{\"{o}}ber S.,   M{\"{u}}ller V.,  2002, ]
  {10.1051/0004-6361:20020303}, 387, 396

\bibitem[\protect\citeauthoryear{Bassett \& Foster}{Bassett \&
  Foster}{2019}]{Bassett2019ProspectsQuantities}
Bassett R.,  Foster C.,  2019, \mndoi [Monthly Notices of the Royal
  Astronomical Society] {10.1093/mnras/stz1440}, 487, 2354

\bibitem[\protect\citeauthoryear{Bianchini, Norris, van~de Ven  \&
  Schinnerer}{Bianchini et~al.}{2015}]{Bianchini2015UnderstandingObservations}
Bianchini P.,  Norris M.~A.,  van~de Ven G.,   Schinnerer E.,  2015, ]
  {10.1093/mnras/stv1651}

\bibitem[\protect\citeauthoryear{Binney}{Binney}{2005}]{Binney2005RotationRevisited}
Binney J.,  2005, MNRAS, 363, 937

\bibitem[\protect\citeauthoryear{Binney \& Tremaine}{Binney \&
  Tremaine}{2008}]{Binney2008GalacticDynamics}
Binney J.,  Tremaine S.,  2008, {Galactic Dynamics}, second edn.
Princeton Series in Astrophysics, Princeton University Press

\bibitem[\protect\citeauthoryear{Blanton et~al.,}{Blanton
  et~al.}{2017}]{Blanton2017SloanUniverse}
Blanton M.~R.,  et~al., 2017, \mndoi [The Astronomical Journal]
  {10.3847/1538-3881/aa7567}, 154, 28

\bibitem[\protect\citeauthoryear{Bruzual \& Charlot}{Bruzual \&
  Charlot}{2003}]{Bruzual2003Stellar2003}
Bruzual G.,  Charlot S.,  2003, ] {10.1046/j.1365-8711.2003.06897.x}, 344, 1000

\bibitem[\protect\citeauthoryear{Bryant et~al.,}{Bryant
  et~al.}{2015}]{Bryant2015TheSelection}
Bryant J.~J.,  et~al., 2015, \mndoi [Monthly Notices of the Royal Astronomical
  Society] {10.1093/mnras/stu2635}, 447, 2857

\bibitem[\protect\citeauthoryear{Bullock, Dekel, Kolatt, Kravtsov, Klypin,
  Porciani  \& Primack}{Bullock et~al.}{2001}]{Bullock2001AHalos}
Bullock J.~S.,  Dekel A.,  Kolatt T.~S.,  Kravtsov A.~V.,  Klypin A.~A.,
  Porciani C.,   Primack J.~R.,  2001, ApJ, 555, 240

\bibitem[\protect\citeauthoryear{Bundy et~al.,}{Bundy
  et~al.}{2015}]{Bundy2015OverviewObservatory}
Bundy K.,  et~al., 2015, {Overview of the SDSS-IV MaNGA survey: Mapping Nearby
  Galaxies at Apache Point Observatory}, \mndoi{10.1088/0004-637X/798/1/7}

\bibitem[\protect\citeauthoryear{Cappellari}{Cappellari}{2017}]{Cappellari2017ImprovingFunctions}
Cappellari M.,  2017, \mndoi [Monthly Notices of the Royal Astronomical
  Society] {10.1093/mnras/stw3020}, 466, 798

\bibitem[\protect\citeauthoryear{Cappellari \& Copin}{Cappellari \&
  Copin}{2003}]{Cappellari2003AdaptiveTessellations}
Cappellari M.,  Copin Y.,  2003, \mndoi [Monthly Notices of the Royal
  Astronomical Society] {10.1046/j.1365-8711.2003.06541.x}, 342, 345

\bibitem[\protect\citeauthoryear{Cappellari, Emsellem, Bacon  \&
  {others}}{Cappellari et~al.}{2007}]{Cappellari2007TheKinematics}
Cappellari M.,  Emsellem E.,  Bacon R.,   {others} 2007, MNRAS, 379, 418

\bibitem[\protect\citeauthoryear{Cappellari, Emsellem, Krajnovic  \&
  {others}}{Cappellari et~al.}{2011}]{Cappellari2011TheRelation}
Cappellari M.,  Emsellem E.,  Krajnovic D.,   {others} 2011, MNRAS, 416, 1680

\bibitem[\protect\citeauthoryear{Chabrier}{Chabrier}{2003}]{Chabrier2003GalacticFunction}
Chabrier G.,  2003, ] {10.1086/376392}, 115, 763

\bibitem[\protect\citeauthoryear{Charlot \& Fall}{Charlot \&
  Fall}{2000}]{Charlot2000AGalaxies}
Charlot S.,  Fall S.~M.,  2000, ] {10.1086/309250}, 539, 718

\bibitem[\protect\citeauthoryear{Cortese, Fogarty, Bekki  \& {others}}{Cortese
  et~al.}{2016}]{Cortese2016TheMorphology}
Cortese L.,  Fogarty L. M.~R.,  Bekki K.,   {others} 2016, MNRAS, 463, 170

\bibitem[\protect\citeauthoryear{Crain et~al.,}{Crain
  et~al.}{2015}]{Crain2015TheVariations}
Crain R.~A.,  et~al., 2015, ] {10.1093/mnras/stv725}, 450, 1937

\bibitem[\protect\citeauthoryear{Croom et~al.,}{Croom
  et~al.}{2012}]{Croom2012TheSpectrograph}
Croom S.~M.,  et~al., 2012, \mndoi [Monthly Notices of the Royal Astronomical
  Society] {10.1111/j.1365-2966.2011.20365.x}, 421, 872

\bibitem[\protect\citeauthoryear{D'Eugenio, Houghton, Davies  \&
  {others}}{D'Eugenio et~al.}{2013}]{DEugenio2013FastZ=0.183}
D'Eugenio F.,  Houghton R. C.~W.,  Davies R.~L.,   {others} 2013, MNRAS, 429,
  1258

\bibitem[\protect\citeauthoryear{Dale, Helou, Magdis, Armus,
  D{\textbackslash}'{\{}{\textbackslash}i{\}}az-Santos  \& Shi}{Dale
  et~al.}{2014}]{Dale2014ANuclei}
Dale D.~A.,  Helou G.,  Magdis G.~E.,  Armus L.,
  D{\textbackslash}'{\{}{\textbackslash}i{\}}az-Santos T.,   Shi Y.,  2014, ]
  {10.1088/0004-637X/784/1/83}, 784, 83

\bibitem[\protect\citeauthoryear{De~Vita, Trenti, Bianchini, Askar, Giersz  \&
  van~de Ven}{De~Vita et~al.}{2017}]{DeVita2017ProspectsSpectroscopy}
De~Vita R.,  Trenti M.,  Bianchini P.,  Askar A.,  Giersz M.,   van~de Ven G.,
  2017, \mndoi [Monthly Notices of the Royal Astronomical Society]
  {10.1093/mnras/stx325}, 467, 4057

\bibitem[\protect\citeauthoryear{Duckworth, Tojeiro  \& Kraljic}{Duckworth
  et~al.}{2020}]{Duckworth2020DecouplingSpin}
Duckworth C.,  Tojeiro R.,   Kraljic K.,  2020, \mndoi [Monthly Notices of the
  Royal Astronomical Society] {10.1093/mnras/stz3575}, 492, 1869

\bibitem[\protect\citeauthoryear{Emsellem, Cappellari, Krajnovic  \&
  {others}}{Emsellem et~al.}{2007}]{Emsellem2007TheGalaxies}
Emsellem E.,  Cappellari M.,  Krajnovic D.,   {others} 2007, MNRAS, 379, 401

\bibitem[\protect\citeauthoryear{Few, Courty, Gibson, Kawata, Calura  \&
  Teyssier}{Few et~al.}{2012}]{Few2012RAMSES-CH:Simulations}
Few C.~G.,  Courty S.,  Gibson B.~K.,  Kawata D.,  Calura F.,   Teyssier R.,
  2012, ] {10.1111/j.1745-3933.2012.01275.x}, 424, L11

\bibitem[\protect\citeauthoryear{Genel et~al.,}{Genel
  et~al.}{2014}]{Genel2014IntroducingTime}
Genel S.,  et~al., 2014, ] {10.1093/mnras/stu1654}, 445, 175

\bibitem[\protect\citeauthoryear{Genel, Fall, Hernquist  \& {others}}{Genel
  et~al.}{2015}]{Genel2015GalacticSequence}
Genel S.,  Fall M.,  Hernquist L.,   {others} 2015, ApJ, 804, 7pp

\bibitem[\protect\citeauthoryear{Graham, Cappellari, Li  \& {others}}{Graham
  et~al.}{2018}]{Graham2018SDSS-IVProperties}
Graham M.~T.,  Cappellari M.,  Li H.,   {others} 2018, MNRAS, sty504

\bibitem[\protect\citeauthoryear{Green et~al.,}{Green
  et~al.}{2018}]{Green2018TheProducts}
Green A.~W.,  et~al., 2018, \mndoi [Monthly Notices of the Royal Astronomical
  Society] {10.1093/mnras/stx3135}, 475, 716

\bibitem[\protect\citeauthoryear{Greene, Leathaud, Emsellem  \&
  {others}}{Greene et~al.}{2018}]{Greene2018SDSS-IVGalaxies}
Greene J.~E.,  Leathaud A.,  Emsellem E.,   {others} 2018, ApJ, 852, 36

\bibitem[\protect\citeauthoryear{Gu{\'{e}}rou et~al.,}{Gu{\'{e}}rou
  et~al.}{2017}]{Guerou2017The0.8}
Gu{\'{e}}rou A.,  et~al., 2017, ] {10.1051/0004-6361/201730905}, 608, A5

\bibitem[\protect\citeauthoryear{Harborne}{Harborne}{2019}]{Harborne2019SimSpin:Simulations}
Harborne K.,  2019, Astrophysics Source Code Library, record ascl:1903.006

\bibitem[\protect\citeauthoryear{Harborne, Power, Robotham, Cortese  \&
  Taranu}{Harborne et~al.}{2019}]{Harborne2019A_R}
Harborne K.~E.,  Power C.,  Robotham A. S.~G.,  Cortese L.,   Taranu D.~S.,
  2019, MNRAS, 483, 249

\bibitem[\protect\citeauthoryear{Jesseit, Cappellari, Naab  \&
  {others}}{Jesseit et~al.}{2009}]{Jesseit2009SpecificParameter}
Jesseit R.,  Cappellari M.,  Naab T.,   {others} 2009, MNRAS, 397, 1202

\bibitem[\protect\citeauthoryear{Lagos, Stevens, Bower  \& {others}}{Lagos
  et~al.}{2018a}]{Lagos2018QuantifyingGalaxies}
Lagos C. d.~P.,  Stevens A. R.~H.,  Bower R.~G.,   {others} 2018a, MNRAS, 473,
  4956

\bibitem[\protect\citeauthoryear{Lagos, Schaye, Bah{\'{e}}, Van~de Sande, Kay,
  Barnes, Davis  \& Vecchia}{Lagos et~al.}{2018b}]{Lagos2018TheGalaxies}
Lagos C. d.~P.,  Schaye J.,  Bah{\'{e}} Y.,  Van~de Sande J.,  Kay S.~T.,
  Barnes D.,  Davis T.~A.,   Vecchia C.~D.,  2018b, \mndoi [Monthly Notices of
  the Royal Astronomical Society] {10.1093/mnras/sty489}, 476, 4327

\bibitem[\protect\citeauthoryear{Lagos, Tobar, Robotham, Obreschkow, Mitchell,
  Power  \& Elahi}{Lagos et~al.}{2018c}]{Lagos2018Shark:Formation}
Lagos C. d.~P.,  Tobar R.~J.,  Robotham A.~S.,  Obreschkow D.,  Mitchell P.~D.,
   Power C.,   Elahi P.~J.,  2018c, \mndoi [Monthly Notices of the Royal
  Astronomical Society] {10.1093/mnras/sty2440}, 481, 3573

\bibitem[\protect\citeauthoryear{McAlpine et~al.,}{McAlpine
  et~al.}{2016}]{McAlpine2016TheCatalogues}
McAlpine S.,  et~al., 2016, \mndoi [Astronomy and Computing]
  {10.1016/j.ascom.2016.02.004}, 15, 72

\bibitem[\protect\citeauthoryear{Moffat}{Moffat}{1969}]{Moffat1969ANASA/ADS}
Moffat A. F.~J.,  1969, Astronomy and Astrophysics, 3, 455

\bibitem[\protect\citeauthoryear{Naab, Oser, Emsellem  \& {others}}{Naab
  et~al.}{2014}]{Naab2014TheRotators}
Naab T.,  Oser L.,  Emsellem E.,   {others} 2014, MNRAS, 444, 3357

\bibitem[\protect\citeauthoryear{Overzier, Lemson, Angulo, Bertin, Blaizot,
  Henriques, Marleau  \& White}{Overzier et~al.}{2013}]{Overzier2013TheLight}
Overzier R.,  Lemson G.,  Angulo R.~E.,  Bertin E.,  Blaizot J.,  Henriques
  B.~M.,  Marleau G.~D.,   White S.~D.,  2013, \mndoi [Monthly Notices of the
  Royal Astronomical Society] {10.1093/mnras/sts076}, 428, 778

\bibitem[\protect\citeauthoryear{Papastergis \& Ponomareva}{Papastergis \&
  Ponomareva}{2017}]{Papastergis2017TestingDwarfs}
Papastergis E.,  Ponomareva A.~A.,  2017, Astronomy and Astrophysics, 601, 6
  pp.

\bibitem[\protect\citeauthoryear{Pedrosa \& Tissera}{Pedrosa \&
  Tissera}{2015}]{Pedrosa2015AngularScenario}
Pedrosa S.~E.,  Tissera P.~B.,  2015, Astronomy and Astrophysics, 584, 8

\bibitem[\protect\citeauthoryear{Pillepich et~al.,}{Pillepich
  et~al.}{2018}]{Pillepich2018SimulatingModel}
Pillepich A.,  et~al., 2018, \mndoi [Monthly Notices of the Royal Astronomical
  Society] {10.1093/mnras/stx2656}, 473, 4077

\bibitem[\protect\citeauthoryear{{Planck Collaboration} et~al.,}{{Planck
  Collaboration} et~al.}{2018}]{PlanckCollaboration2018PlanckParameters}
{Planck Collaboration} et~al., 2018, arXiv e-prints, p. arXiv:1807.06209

\bibitem[\protect\citeauthoryear{Robotham, Taranu, Tobar  \& {others}}{Robotham
  et~al.}{2017}]{Robotham2017ProFitImages}
Robotham A. S.~G.,  Taranu D.~S.,  Tobar R.,   {others} 2017, MNRAS, 466, 1513

\bibitem[\protect\citeauthoryear{Robotham, Bellstedt, Lagos, Thorne, Davies,
  Driver  \& Bravo}{Robotham et~al.}{2020}]{Robotham2020ProSpect:Histories}
Robotham A. S.~G.,  Bellstedt S.,  Lagos C. d.~P.,  Thorne J.~E.,  Davies
  L.~J.,  Driver S.~P.,   Bravo M.,  2020

\bibitem[\protect\citeauthoryear{Schaye et~al.,}{Schaye
  et~al.}{2015}]{Schaye2015TheEnvironments}
Schaye J.,  et~al., 2015, \mndoi [Monthly Notices of the Royal Astronomical
  Society] {10.1093/mnras/stu2058}, 446, 521

\bibitem[\protect\citeauthoryear{Springel}{Springel}{2005}]{Springel2005TheGadget-2}
Springel V.,  2005, MNRAS, 364, 1105

\bibitem[\protect\citeauthoryear{Springel \& Hernquist}{Springel \&
  Hernquist}{2003}]{Springel2003CosmologicalFormation}
Springel V.,  Hernquist L.,  2003, ] {10.1046/j.1365-8711.2003.06206.x}, 339,
  289

\bibitem[\protect\citeauthoryear{Springel, White, Jenkins  \&
  {others}}{Springel et~al.}{2005}]{Springel2005SimulationsQuasars}
Springel V.,  White S. D.~M.,  Jenkins A.,   {others} 2005, Nature, 435, 629

\bibitem[\protect\citeauthoryear{Teklu, Remus, Dolag  \& {others}}{Teklu
  et~al.}{2015}]{Teklu2015ConnectingMorphology}
Teklu A.~F.,  Remus R.-S.,  Dolag K.,   {others} 2015, ApJ, 812, 24 pp.

\bibitem[\protect\citeauthoryear{Torrey et~al.,}{Torrey
  et~al.}{2015}]{Torrey2015SyntheticSimulation}
Torrey P.,  et~al., 2015, \mndoi [Monthly Notices of the Royal Astronomical
  Society] {10.1093/mnras/stu2592}, 447, 2753

\bibitem[\protect\citeauthoryear{Vazdekis, Koleva, Ricciardelli, R{\"{o}}ck  \&
  Falc{\'{o}}n-Barroso}{Vazdekis
  et~al.}{2016}]{Vazdekis2016UV-extendedGalaxies}
Vazdekis A.,  Koleva M.,  Ricciardelli E.,  R{\"{o}}ck B.,
  Falc{\'{o}}n-Barroso J.,  2016, ] {10.1093/mnras/stw2231}, 463, 3409

\bibitem[\protect\citeauthoryear{Vogelsberger, Genel, Springel  \&
  {others}}{Vogelsberger et~al.}{2014}]{Vogelsberger2014PropertiesSimulation}
Vogelsberger M.,  Genel S.,  Springel V.,   {others} 2014, Nature, 509, 177

\bibitem[\protect\citeauthoryear{Yurin \& Springel}{Yurin \&
  Springel}{2014}]{Yurin2014AnEquilibrium}
Yurin D.,  Springel V.,  2014, MNRAS, 444, 62

\bibitem[\protect\citeauthoryear{van~de Sande, Bland-Hawthorn, Brough  \&
  {others}}{van~de Sande et~al.}{2017a}]{vandeSande2017TheSurveys}
van~de Sande J.,  Bland-Hawthorn J.,  Brough S.,   {others} 2017a, MNRAS, 472,
  1272

\bibitem[\protect\citeauthoryear{van~de Sande, Bland-Hawthorn, Fogarty  \&
  {others}}{van~de Sande et~al.}{2017b}]{vandeSande2017TheKinematics}
van~de Sande J.,  Bland-Hawthorn J.,  Fogarty L. M.~R.,   {others} 2017b, ApJ,
  835, 35pp

\bibitem[\protect\citeauthoryear{van~de Sande et~al.,}{van~de Sande
  et~al.}{2018}]{vandeSande2018TheSimulations}
van~de Sande J.,  et~al., 2018, preprint, p. arXiv:1810.10542

\makeatother
\end{thebibliography}

\end{document}